\documentclass[aps,prl,superscriptaddress,twocolumn,showpacs,preprintnumbers,nofootinbib]{revtex4-1}
\usepackage{amsfonts}
\usepackage{amssymb,amsmath}
\usepackage{mathrsfs}
\usepackage{amsthm}
\usepackage{newlfont}
\usepackage{graphicx}

\newcommand{\usb}{\affiliation{Departamento de F\'{\i}sica, Secci\'{o}n de Fen\'{o}menos \'{O}pticos, Universidad Sim\'{o}n Bol\'{\i}var,Apartado Postal 89000, Caracas 1080-A, Venezuela.}}
\newcommand{\ivic}{\affiliation{Centro de F\'{\i}sica, Instituto Venezolano de Investigaciones Cient\'{\i}ficas, Apartado 20632 Caracas 1020-A, Venezuela.}}

\begin{document}
\begin{flushright} ${}$\\[-40pt] $\scriptstyle \mathrm SB/F/431-13$ \\[0pt]
\end{flushright}
\title{Violation of the center of mass theorem for systems with electromagnetic interaction}
\author{Rodrigo Medina}\email[]{rmedina@ivic.gob.ve}\ivic
\author{J Stephany}\email[]{stephany@usb.ve}\usb
\pacs{45.20.df}
\date{\today}

\begin{abstract}
In this letter we show that for isolated systems for which the energy current density is not equal to the momentum density, that means for systems with non-symmetric energy-momentum, the usual statements of the center of mass motion theorem are not valid. We also discuss the modified version of the theorem that is  always valid. Then we present a simple and exactly workable example of an
electromagnetic interacting system which illustrates the violation of those standard versions of the center of mass motion theorem. We show  that for this system  Minkowski's density
of linear momentum which yields a non-symmetric electromagnetic energy-momentum tensor, but not Abraham's, is compatible with total momentum
conservation. 
\end{abstract}

\maketitle

\bigskip

\section{Introduction}
There is a widespread belief in the unrestricted validity of the so called
center of mass motion theorem. There are two versions of this theorem. The
first (CMMT-1) states that the center of mass of an isolated system moves
with constant velocity. The second (CMMT-2) states that the total momentum of
an isolated system equals the mass (energy divided by $c^2$) times the
velocity of the center of mass. The second version implies the first.
In particular both versions of the theorem have been used many times in the century long controversy \cite{Minkowski1909,Abraham1910} about Abraham's and Minkowski's proposals for the electromagnetic momentum in a dielectric, mostly to support Abraham's point of view \cite{Balazs1953,Leonhart2006,Pfeifer2007,Barnett2010,MilonniBoyd2010,BarnettLoudon2010,Cho2013}\footnote{The literature on this theme is too vast to be properly handled here, so we limit our references to few key articles and a couple of review papers to which we refer the reader for further information.}.
Actually, such a theorem  is a consequence of the conservation of orbital angular momentum, which in general is not true. What is always conserved for an isolated relativistic  system is the total angular momentum which also includes spin.
The theorem does not hold when the energy-momentum tensor  that defines the orbital angular momentum is not symmetric. In such a case the density of linear momentum $c^{-1}T^{i0}$ is not proportional to the current density of energy $cT^{0i}$, and the velocity of the center of mass which depends on the latter is not constant. To be more specific, let us consider a localized system upon which no forces are exerted but with a non-symmetric energy-momentum tensor. This may occur for an isolated system if there is a local spin density $S^{\mu\nu\alpha}$ \cite{Papapetrou1949,MandS2014}. The total energy $U=\int T^{00}dV$ and the total momentum
$p^i = c^{-1}\int T^{i0}dV$ are conserved. The current of the orbital angular momentum is
\begin{equation}
\label{orbital}
L^{\mu\nu\alpha} = x^\mu T^{\nu\alpha}-x^\nu T^{\mu\alpha} \ .
\end{equation}
Since no forces act on the isolated system, 
for any sensible definition of the energy-momentum tensor,
$\partial_\nu T^{\mu\nu}=0$, and therefore we have
\begin{equation}
\label{orbital-eq}
\partial_\alpha L^{\mu\nu\alpha} = T^{\nu\mu}-T^{\mu\nu} \ \ .
\end{equation}
and
\begin{equation}
\partial_\alpha S^{\mu\nu\alpha} = -T^{\nu\mu}+T^{\mu\nu} \ \ .
\end{equation}
When integrated over the volume this second equation relates the torques on the system with the variation of spin. For a magnet the spatial part of the spin density is related to magnetization. When the total magnetic moment is different of zero, it is these equations which describe the Einstein-de Haas effect where spin is converted in orbital angular momentum. Einstein-de Haas effect is  used routinely to measure the gyromagnetic radio \cite{BarKen1952} and provides an example where the total energy-momentum tensor cannot be symmetric. 
Define the center of mass by
\begin{equation}
\label{CenterMass}
X^i = \frac{1}{U}\int x^i T^{00}\,dV\ \ .
\end{equation}
Then the temporal component of orbital angular momentum is
\begin{equation}
\label{Li0}
L^{i0} = c^{-1}\int (x^i T^{00} - x^0 T^{i0})dV
=\frac{U}{c}X^i-tcp^i\ \ .
\end{equation}
From (\ref{orbital-eq}) we get
\begin{equation}
\dot{L}^{i0} = \int(T^{0i}-T^{i0})dV
=\int T^{0i}\,dV -cp^i \ \ .
\end{equation}
Using (\ref{Li0}), $\dot{U}=0$ and $\dot{p}^i=0$, one obtains
the velocity of the center of mass
\begin{equation}
\label{CenterMass-velocity}
\dot{X}^i = \frac{c}{U}\int T^{0i}\,dV\ \ .
\end{equation}
Note that what appears in (\ref{CenterMass-velocity}) is the energy current
density, not the momentum density.
This third form of the theorem about the motion of center of mass (CMMT-3) is always
valid for isolated systems.
When the energy-momentum tensor is symmetric, $\dot{X}^i = c^2 p^i/U $
and the center of mass velocity is constant, but in general it is not so.

This discussion is relevant to the  Abraham and Minkowski debate on the energy-momentum tensor $T^{\mu\nu}_F $of the electromagnetic field in a medium. There has been agreement on the energy current
density which is Poynting's vector 
\begin{equation}
\mathbf{S} = c T^{0i}_F \hat{\mathbf{e}}_i =
\frac{c}{4\pi}\mathbf{E}\times\mathbf{H}\ \ ,
\end{equation}
but Minkowski's  density of linear momentum,
\begin{equation}
\label{momentum-density}
\mathbf{g}_{\mathrm{Min}} = c^{-1} T^{i0}_F \hat{\mathbf{e}}_i =
\frac{1}{4\pi c}\mathbf{D}\times\mathbf{B}\ \ ,
\end{equation}
has been challenged for many years by the alternative suggestion of Abraham,
$\mathbf{g}_{\mathrm{Abr}}=c^{-2}\mathbf{S}$, which yields a symmetric
 energy-momentum tensor.
Minkowski proposal has been supported for example by arguments using Doppler
effect and the absorption of light within a dielectric.
On the other hand Abraham's momentum density is in general consistent with CMMT-2. 
Many other arguments on Abraham's side, beginning with Balazs work
\cite{Balazs1953}, rely on the use of CMMT-1 or CMMT-2, sometimes in a
classical formulation and sometimes expressed in terms of photons
\cite{Pfeifer2007,MilonniBoyd2010}. The main objective of this letter is to
show the flaw of this line of reasoning. 
Below we present a simple example which illustrates the violation of the first
two versions of the center of mass theorem in the presence of electromagnetic
fields without relying in any specific definition of the energy-momentum
tensor. We then  show that  use of Minkowski's density of linear momentum is
compatible with momentum conservation. Our example is related to  Shockley's and James configuration \cite{SJ1967} which also shows the inconsistency of CMMT-2 and standard electromagnetism and led them to postulate the existence of a hidden momentum. The system discussed here  has the advantage of being simpler and fully computable. We should stress that in view of our discussion above the hypothesized hidden momentum is completely unnecessary.

\section{Force on a magnet}
Consider a non-charged solid body at rest made of insulating material
with dielectric constant $\epsilon=1$ and magnetization $\mathbf{M}$.
In these conditions $\rho=0$, $\mathbf{j}=0$ and $\mathbf{P}=0$. Assuming 
that the force on the body is due to the Lorentz force on the
microscopic charges and currents in  standard
electrodynamics it is well-known \cite{JacJ1998}
that the  force on a magnetic dipole $\mathbf{m}$ is
\begin{equation}
\label{force-dipole}
\mathbf{F}=\nabla(\mathbf{m}\cdot\mathbf{B})
\end{equation}
and the corresponding power is $-\mathbf{m}\cdot\dot{\mathbf{B}}$. They are independent of the electric field $\mathbf{E}$ in the rest reference frame. Since each element of the body can be considered as a magnetic dipole with moment $d\mathbf{m}=\mathbf{M}dV$, the four-force density due to
electromagnetic interactions is
\begin{equation} 
\label{force-density}
f^\mu =M_k\partial^\mu B_k\ \ .
\end{equation}
This expression may be written in an explicit covariant form by including the terms depending on the polarization vector which appear in a moving frame. Note
that this force density is not the same than the force density on
the magnetic current density $\mathbf{j}_M = c\nabla\times\mathbf{M}$ although it may be shown that the total force on the body computed with any one of them yields the same result.
Other expressions or contributions to the force density have been proposed \cite{Pfeifer2007,MilonniBoyd2010} , in
particular  Einstein-Laub force,  Abraham force and the contribution of hidden momentum.
These forces either are $\mathbf{E}$-dependent, quadratic in the magnetization or rely in a  non-standard treatment of the dynamics. (See also the resource letter  \cite{GriD20121} for more references).

In general the magnetic induction field is composed by the field produced
by the magnetization itself and an external field
$\mathbf{B}=\mathbf{B}_s+\mathbf{B}_e$. When $\mathbf{M}$ is changing smoothly,  the self-interaction force is compensated by the stress that develops in the material. In order to compute
the total force on the body its is sufficient to consider the external
field $\mathbf{B}_e$.

\section{The magnet-charge system}
\begin{figure}
\centering
\includegraphics[scale=0.8]{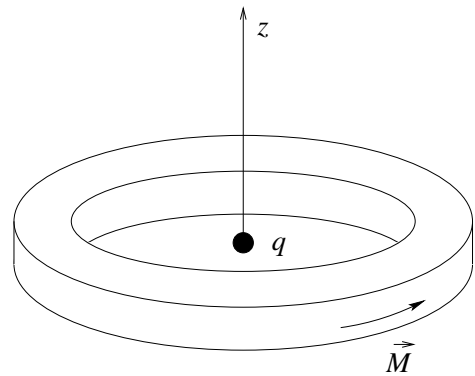}
\caption{Magnetized ring}
\label{washer}
\end{figure}

Consider a ring with the shape of a washer made of a ferromagnetic insulator
(see Fig.~\ref{washer}). The axis of the ring is along the $z$-axis and the
center is at the origin. For simplicity we take $\epsilon=1$. The ring is magnetized  around its axis with  magnetization $\mathbf{M}$ of uniform magnitude.
In this situation $\mathbf{B}=4\pi\mathbf{M}$ inside the ring whereas
 $\mathbf{H}=0$ everywhere.
Suppose that at the center of the ring there is a particle of  charge $q$. Since the ring is an insulator the electrostatic field of the charge penetrates the magnet. The center of mass of the system (including the electromagnetic contribution to the energy) is at rest at the center of the ring.
Abraham's momentum density vanishes since $\mathbf{H}=0$ so that if one takes it to define the momentum of the electromagnetic field,  the total momentum of the system including the electromagnetic field is zero. This is consistent with CMMT-2 only if the center of mass remains stationary. On the other hand, taking Minkowski's definition there is a finite momentum in the $z$ direction. The momentum of the electromagnetic field for this second alternative can be easily calculated. Using cylindrical
coordinates $(\rho, \theta, z)$, $\mathbf{M}=M\hat{\mathbf{{\theta}}}$ and
 $\mathbf{r}=\rho\hat{\mathbf{{\rho}}}+ z\hat{\mathbf{z}}$. The momentum density inside the ring is
\begin{equation}
\mathbf{g}_{\mathrm{Min}}=\frac{1}{4\pi c}\mathbf{D}\times\mathbf{B}=
\frac{q M}{c r^3}(-z\hat{\mathbf{\rho}}+\rho \hat{\mathbf{z}})\ \ .
\end{equation}
Integrating over the volume
\begin{equation}
\label{momentum}
\mathbf{P}_{\mathrm{Min}} = \frac{q M}{c}\int dV\, \frac{\rho}{r^3}\,\hat{\mathbf{z}}\ \ .
\end{equation}
The integral on the right side of this equation may be computed in terms of the geometrical parameters of the ring. Accepting Minkowski's definition for the momentum density violates CMMT-2 in this configuration consistently with our discussion at the beginning. 

To determine which of the two proposals for the momentum density is more useful in the description of this system, let us examine what happens when the fields  change. For example consider the scenario in
which the magnet demagnetizes ({\it e.~g.~}by including in the magnet a device
which allows  its temperature to rise above the Curie point). The magnetic
induction field $\mathbf{B}$ changes and therefore an induced electric field
appears. It is always possible to split the electric field in a static field,
$\mathbf{E}_S$, $\nabla\cdot\mathbf{E}_S=4\pi\rho$, $\nabla\times\mathbf{E}_S=0$
and an induced field $\mathbf{E}_{\mathrm{I}}$, $\nabla\cdot\mathbf{E}_{\mathrm{I}}=0$,
$\nabla\times\mathbf{E}_{\mathrm{I}}=-c^{-1}\dot{\mathbf{B}}$.  It is easily  shown that
\begin{equation}
\mathbf{E}_{\mathrm{I}} = -\frac{1}{4\pi c}\int dV\,\frac{\partial\mathbf{B}}{\partial t}
\times \frac{\hat{\mathbf{r}}}{r^2}\ \ .
\end{equation}
 Neglecting the radiation field $\mathbf{B}=4\pi\mathbf{M}$. Then at the origin
the induced field is
\begin{equation}
\mathbf{E}_{\mathrm{I}}(0)=-\frac{\dot{M}}{c}\int dV\,\frac{\rho}{r^3}\,\hat{\mathbf{z}}\ \ .
\end{equation}
If the demagnetization process is fast enough, the force on the charge
is $\mathbf{F}=q\mathbf{E}_{\mathrm{I}}(0)$, and the impulse  on the (dressed) particle may also be computed
\begin{equation}
\mathbf{I}_{\mathrm{Charge}}=\int dt\,\mathbf{F}=-\frac{q}{c}\int dt\,\dot{M}\,
\int dV\,\frac{\rho}{r^3}\,\hat{\mathbf{z}}\ \ .
\end{equation}
This is exactly the momentum calculated in (\ref{momentum}) using Minkowski's
momentum density which   is then shown to be  consistent with momentum conservation.
Note that, since no $\mathbf{B}$ field is produced by the charge, there is no
force acting on the magnet which therefore remains at rest. Since the center of
mass of the magnet does not move, the center of mass of the system clearly
moves in the direction in which the particle moves, and so CMMT-1 does not hold in this case.

Next we investigate  the origin of the momentum that in Minkowski's formulation is stored in the electromagnetic field. Let us start with  the charge  far away and analyze how, as it is brought to the center of the ring, the field momentum builds up. Since the magnet does not produce any electromagnetic
field outside itself there is no force upon the charge. There is, though, a
force acting on the magnet produced by the magnetic field $\mathbf{B}$ generated as
the charge moves. To keep the magnet in place an opposite force must be
applied to it. The impulse generated by this force is precisely the stored momentum in the electromagnetic field. For a charge moving along the $z$-axis the calculation can be done easily.

Call $\mathbf{x}$ the position of the charge, $\mathbf{x}=\zeta\hat{\mathbf{z}}$. The velocity
is $\mathbf{v}=\dot{\zeta}\hat{\mathbf{z}}$.  For $v\ll c$ the magnetic field produced by
the moving charge is
\begin{equation}
\mathbf{B}=\frac{q}{c}\frac{\mathbf{v}\times\mathbf{r}^\prime}{{r^\prime}^3}=
\frac{q\dot{\zeta}\rho\hat{\mathbf{\theta}}}{c[\rho^2+(z-\zeta)^2]^{3/2}} \ \ ,
\end{equation}
where ${\mathbf{r}}^\prime=\mathbf{r}-\mathbf{x}=\rho\hat{\rho}+(z-\zeta)\hat{\mathbf{z}}$.

Using (\ref{force-density}) the force density on the magnet is
\begin{equation}
\mathbf{f}=M_k\nabla B_k=\frac{q}{c}\dot{\zeta}M\left[\hat{\mathbf{\rho}}
\frac{\partial}{\partial\rho}+\hat{\mathbf{z}}\frac{\partial}{\partial z}\right]
\frac{\rho}{{r^\prime}^3}\ \ .
\end{equation}
The force is obtained by integrating over the volume
\begin{eqnarray}
\mathbf{F} = \int dV\,\mathbf{f}&=& \frac{q}{c}\dot{\zeta}M\int dV\,\frac{\partial
}{\partial z}\frac{\rho}{{r^\prime}^3}\,\hat{\mathbf{z}}\\
&=& -\frac{q}{c}M\frac{d}{dt}\int dV\,\frac{\rho}{{r^\prime}^3}\,\hat{\mathbf{z}}\ \ .
\end{eqnarray}
We have used $\partial_z =-\partial_\zeta$. The impulse on the magnet is then
\begin{equation}
\mathbf{I}_{\mathrm{Magnet}}=\int dt\,\mathbf{F}=-\frac{q}{c}M\int dV\,\frac{\rho}{r^3}\,\hat{\mathbf{z}}
\ \ .
\end{equation}
The momentum stored by  the field is the opposite of this quantity and again agrees with (\ref{momentum}).

Also in this situation CMMT-1 does not hold. If no external
force is applied to keep the magnet in place, the charge would move with constant velocity but the magnet would be accelerated in the negative direction of the $z$ axis as the center of mass would be too.

\section{Conclusion}
The example presented in this letter display in a very simple setup the
violation of the center of mass theorem for a system with electromagnetic
interactions. The discussion presented depends only on the Maxwell equations
and uses the standard electromagnetic force in the mechanical equations.
 The  impulse computed in each situation,  on the charge or on the magnet,
does not depend on the definition of the momentum density or any other
component of the energy momentum-tensor of the field.
Nevertheless when confronted with  the alternative definitions for the momentum
density of the field, the results single out Minkowski's proposal as the one
which allows to satisfy linear momentum conservation. Or equivalently, the one
which is consistent with  the action-reaction law between the electromagnetic
field and matter. 

We also note that in this example the electromagnetic field acts as a buffer
for the linear momentum, spoiling the validity of Newton's third law for the
matter sector of the system. In the first situation the linear momentum has
been stored in the field although everything in the system is at rest. When the
momentum is released the magnet feels no reaction although the particle is
accelerated. In the second situation no force acts on the particle while it moves to be 
placed at the center of the ring. Only the magnet is subjected to the action of
an electromagnetic force, which should then be compensated by an external force
in order to establish the configuration. In the process the field accumulates
linear momentum. 

The situation discussed gives a realization of a system which stores linear momentum and has all its components at rest, giving a way out to the apparent paradoxes introduced by  the wrong use of the center of mass theorem.

Although in this letter we do not enter into the details of how to define
convenient energy-momentum tensors for  field and  matter, something that will
be done elsewhere \cite{MedandSb}, our computation already shows that the
total energy momentum-tensor of this system should be non-symmetric. This is
consistent with the fact that both the field and the matter distribution, may
have a non-vanishing spin density current \cite{Papapetrou1949}, which is what
allows the violation of the first two versions of the center of mass theorem
discussed at the beginning.

Taylor \cite{Taylor1965} has treated a related example, with toroidal currents
instead of a magnet. It is interesting to have a view on the differences with our discussion in the starting situation with the charge at rest in the center. In this  case the proper energy-momentum tensor is known to be the standard
tensor defined for the vacuum , which is symmetric, $\mathbf{S}=c^2\mathbf{g}=
c\mathbf{E}\times\mathbf{B}/4\pi$, and therefore CMMT-2 holds. If one considers
a  free current distribution of value $\mathbf{j}=c\nabla\times\mathbf{M}$, with $\mathbf{M}$ the magnetization of our previous example, the magnetic 
induction field and the induced electric field are of course the same as in that case and the
electromagnetic momentum density is equal to the one obtained using Minkowski's expression in that situation. But now unlike in the magnet configuration the Poynting vector does not
vanish. The force exert on the currents by $\mathbf{E}$ is still zero, but the power density
$\mathbf{E}\cdot\mathbf{j}$ is not. In the equivalent situation to our magnet, inside the structure that supports the currents the energy is increasing in the top  and decreasing in the base. The total power does vanish. The supporting structure is only
apparently at rest, its center of mass moves because there is transport of energy by Poynting's vector. Now CMMT-2 imposes that the velocity of the center of mass equals
the stored momentum divided by the total energy. This makes this configuration ephemeral. Even disregarding Joule effect it could last only until the energy storage in the base is depleted. Regretfully the analysis presented in Taylor's paper is not correct because it does not take into account adequately the electromagnetic momentum. The energy transport mechanism just described is in contrast to what occurs with  the magnet where the power density and the Poynting vector are zero. For this reason the center of mass of the magnet remains always at rest.  

The original Abraham-Minkowski controversy has been considered conceptually
solved by some authors on the ground that the division of the total
energy-momentum tensor into electromagnetic and material components is
arbitrary and hence the Minkowski electromagnetic energy-momentum tensor,
like the Abraham tensor, has a material counterpart in such a way that the sum
of these components yields the same total energy-momentum tensor
\cite{Pfeifer2007}.  Other authors suggest to consider both options correct but
applicable in different contexts (see \cite{Barnett2010} for a recent example).
Nevertheless none of these points of view have been completely adopted by the
community and  discussion continues to take place. We think that the
calculation presented in this letter clearly chooses Minkowski's linear
momentum density (\ref{momentum-density}) as a better tool for the description
of the system. From our point of view it appears to be no satisfactory way to define a modified
energy-momentum tensor of the matter distribution which will store the needed linear
momentum while the magnet remains at rest. One effort in this direction which originated precisely in trying to make things compatible with the (non necessarily true) CMMT-1 and CMMT-2 is the use of a hidden momentum \cite{SJ1967,GriD20121}. An analysis based on it may be used to force these theorems to hold. We thing that the mechanism through which the hypothetical hidden momentum may be first stored and then transfered to matter is much more obscure and controversial that anything in our discussion. We also think that the use of Einstein-Laub force which also allows to rescue CMMT-1 and CMMT-2 in this case is a risky departure of the experimentally well supported Maxwell-Lorentz framework.   But of course it should be the task of experiments to vindicate or to completely  rule  out these alternatives. On relation with this it should be noted that the magnitude of the stored momentum in the device is very small even in the most optimistic laboratory conditions but not completely out of the possibilities of 
measurement now or in a near future. 


\begin{thebibliography}{long}
\bibitem{Minkowski1909} H.~Minkowski, Nachr. Ges. Wiss. Gottingen, 53 (1909).
\bibitem{Abraham1910} M.~Abraham, Rend. Circ. Mat. Palermo \textbf{30}, 33 (1910).
\bibitem{Balazs1953} N.~L.~Balazs, Phys.~Rev. \textbf{91}, 408--411 (1953).
\bibitem{Leonhart2006} U.~Leonhart, Nature \textbf{444}, 823--824 (2006).
\bibitem{Pfeifer2007} Robert N.~C.~Pfeifer {\it et al}, Rev.~Mod.~Phys.
\textbf{79}, 1197--1216 (2007).
\bibitem{Barnett2010} Stephen M.~Barnett, Phys.~Rev.~Lett. A \textbf{104}, 070401 (2010).
\bibitem{MilonniBoyd2010} Peter W.~Milonni and Robert W.~Boyd, Advances in
Optics and Photonics \textbf{2}, 519--553 (2010).
\bibitem{BarnettLoudon2010} Stephen M.~Barnett and Rodney Loudon, Phil.~
Trans.~R.~Soc. A \textbf{368}, 927--939 (2010).
\bibitem{Cho2013} Adrian Cho, Science \textbf{327} 1067 (2013).
\bibitem{Papapetrou1949} A.~Papapetrou, Phil. Mag. \textbf{40}, 937--946 (1949).
\bibitem{MandS2014} Rodrigo Medina and J.~Stephany, In preparation. 
\bibitem{BarKen1952}S.J.~Barnett and G.S.~Kenny, Phys.~Rev. \textbf{91}, 408--411 (1953).
\bibitem{SJ1967} W.~Shockley and R.P.~James, Phys.~Rev.~Lett. A \textbf{18}, 876 (1967).
\bibitem{JacJ1998} J.~D.~Jackson, {\it Classical Electrodynamics 3rd ed.}, John Wiley\&Sons, New York, 1998, p. 189, p. 213.
\bibitem{GriD20121} D.J.~Griffiths, Am. J. Phys. \textbf{80}, 7--18 (2012).
\bibitem{MedandSb} Rodrigo Medina and J.~Stephany, {\it The force density and the kinetic energy-momentum tensor of electromagnetic fields in matter}, Preprint.
\bibitem{Taylor1965} T.~T.~Taylor, Phys.~Rev.~\textbf{137}, B467--B471 (1965).
\end{thebibliography}
\end{document}